\documentclass[aps,pra,reprint,amsmath,amssymb,showpacs]{revtex4-1}
\usepackage{graphicx}
\usepackage[colorlinks,allcolors=blue]{hyperref}
\usepackage{color}
\usepackage{soul}
\usepackage{ulem}

\definecolor{tmhcolor}{rgb}{0.9,0.3,0.1}
\newcommand{\tmh}[1]{{\color{tmhcolor}{#1}}}

\definecolor{psmcolor}{rgb}{0.1,0.3,0.9}

\begin{document}

\title{Suppression of Qubit Crosstalk in a Tunable Coupling Superconducting Circuit}
\author{Pranav Mundada}\thanks{These authors contributed equally to this work.}
\author{Gengyan Zhang}\thanks{These authors contributed equally to this work.}
\author{Thomas Hazard}\thanks{These authors contributed equally to this work.}
\author{Andrew Houck}\email{aahouck@princeton.edu}
\affiliation{Department of Electrical Engineering, Princeton University, Princeton, New Jersey 08540, USA}
\date{\today}

\begin{abstract}
Parasitic crosstalk in superconducting quantum devices is a leading limitation for quantum gates. We demonstrate the suppression of static \textit{ZZ} crosstalk in a two-qubit, two-coupler superconducting circuit, where the frequency of a tunable coupler can be adjusted such that the \textit{ZZ} interaction from each coupler destructively interfere.  We verify the crosstalk elimination with simultaneous randomized benchmarking, and use a parametrically activated iSWAP interaction to achieve a Bell state preparation fidelity of 98.5\% and a $\sqrt{\mathrm{iSWAP}}$ gate fidelity of 94.8\%  obtained via quantum process tomography.
\end{abstract}
%Using a cross-Ramsey experiment, we measure the strength of this crosstalk as a function of coupler frequency to find the null point, and


\maketitle

Circuit quantum electrodynamics (cQED) \cite{blais2004}, which uses superconducting circuits as its building blocks, has become a promising candidate and testbed for implementing quantum computation. Remarkable research progress has been made in integrating more qubits, resonators and other circuit elements in order to build increasingly computationally powerful devices~\cite{kandala2017,boixo2018}. As the number of circuit elements and control signals scales up in a cQED device, undesirable responses to external controls and unwanted interactions between subsystems lead to crosstalk that must be carefully calibrated and eliminated to ensure optimal device performance~\cite{takita2017,neill2018}. The trade-off between strong qubit-qubit interaction (for fast gates) and low crosstalk poses constraints on the device design and pulse schemes~\cite{gambetta2012,sheldon2016}.

In a cQED system where multiple transmon qubits~\cite{koch2007} are coupled via bus cavities~\cite{majer2007}, the couplings between their higher energy levels give rise to cross-Kerr interactions that can be described by $\zeta a_i^\dagger a_i a_j^\dagger a_j$~\cite{dicarlo2009}, where $ a\ (a^\dagger)$ is the annihilation (creation) operator for the qubit modes, and $\zeta$ corresponds to the frequency shift of qubit $i$ depending on the state of qubit $j$ (and vice versa). This type of static \textit{ZZ} crosstalk causes dephasing in the qubits and degrades device performance if $\zeta$ is comparable to the qubit decoherence rate. In particular, it limits the fidelity of \textit{XX}-type parity measurements in several quantum error correction schemes~\cite{takita2016} and the lifetime of logical qubits containing \textit{XX}-type stabilizers~\cite{takita2017}. Theoretical and experimental studies have shown that \textit{ZZ} crosstalk has become the limiting factor for gate fidelity as qubit coherence times keep improving in state-of-the-art devices~\cite{mckay2017}.

\begin{figure*}
    \centering
    \includegraphics{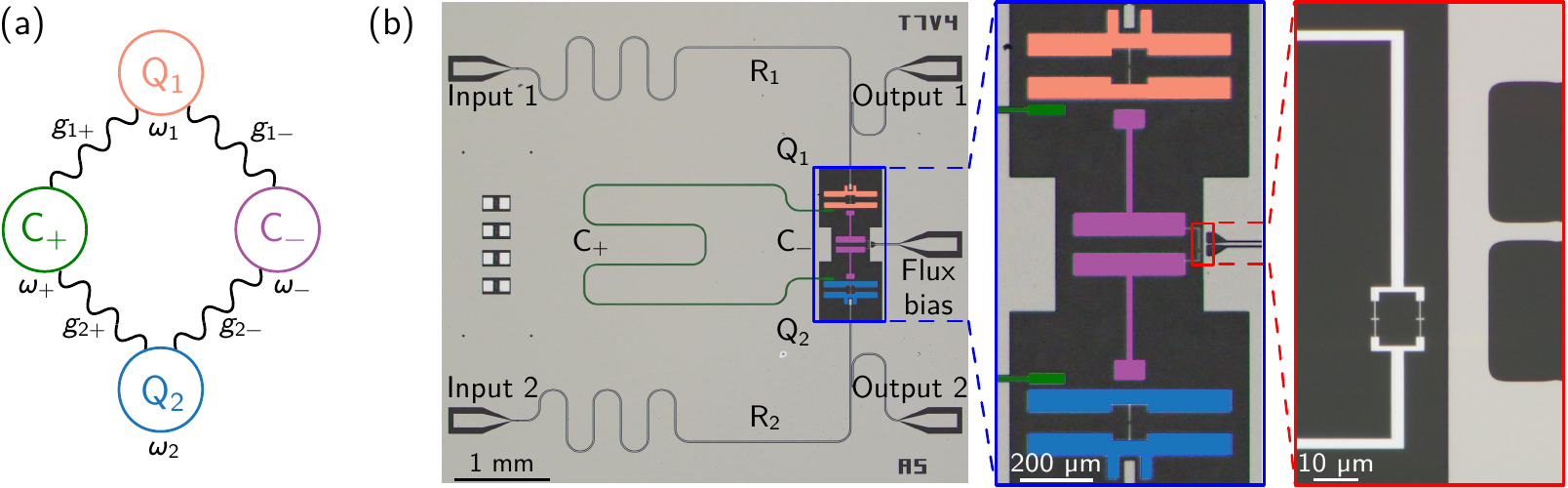}
    \caption{Tunable coupling device for suppression of \textit{ZZ} crosstalk. (a) Conceptual schematic -- we utilize quantum interference between two couplers to achieve zero \textit{ZZ} crosstalk. The device consists of two qubits ($\text{Q}_1,\,\text{Q}_2$) coupled via a bus cavity ($\text{C}_+$) and a tunable coupler ($\text{C}_-$). In order to achieve zero \textit{ZZ} crosstalk and large coupling between the qubits, it is important to have the two qubits in the straddling regime and dispersively coupled to the two couplers \cite{supplement}.   (b) Device A micrograph --  the two qubits are fixed frequency transmons with separate readout resonators (R$_1$ and R$_2$).  A $\lambda/2$ coplanar waveguide resonator is used as the bus cavity.  The tunable coupler consists of a SQUID loop capacitively coupled to each of the two qubits, and its frequency is set by the current through the on-chip bias line (depicted in the rightmost panel).}
    \label{fig:Device}
\end{figure*}

In this work, we utilize quantum interference in a tunable coupling device to demonstrate the suppression of static \textit{ZZ} crosstalk. By introducing a tunable coupler in addition to the bus cavity, shown schematically in ~\autoref{fig:Device}(a), $\zeta$ can be tuned to zero and that an efficient two-qubit gate can be implemented with $\zeta = 0$. Nulling of the always-on \textit{ZZ} interaction is verified by simultaneous randomized benchmarking (RB). Parametrically activated entangling gates, which have been widely employed in superconducting circuits~\cite{mckay2016,niskanen2007,reagor2018,naik2017,roth2017,ganzhorn2018}, can be easily implemented in this architecture. While modulating the coupler frequency at the $\zeta = 0$ point, we characterize the gate fidelity with quantum process tomography and prepare a Bell state with a concurrence of $\mathcal{C}=0.99$.

%The tunability of the two-qubit coupling strength enables us to implement an iSWAP gate using parametric modulation.  \psm{Maybe mention that we use QPT to measure the fidelity of iSWAP} \tmhout{We observe that adequate thermalization of the tunable coupler is crucial for high fidelity iSWAP gate. The dependence of parametric gates on thermal population in the tunable coupler has not been investigated before. We provide a preliminary examination of this in the supplementary materials. Device B is designed to achieve zero ZZ at higher tunable coupler frequency to reduce it's thermal population.}

The device, shown in \autoref{fig:Device}(b), consists of two computational qubits ($\text{Q}_1,\,\text{Q}_2)$ coupled via a tunable coupler ($\text{C}_-$) and a bus cavity ($\text{C}_+$). The Hamiltonian for the device is
\begin{equation}\label{eq:TunableCoupler}
\begin{aligned}
H/\hbar &= \sum_{i=1,2,\pm}{\left(\omega_i a_i^\dagger a_i - \frac{\alpha_i}{2}a_i^\dagger a_i^\dagger a_i a_i\right)}\\
&\quad + \sum_{\substack{i=1,2\\j=\pm}}{g_{ij}(a_i^\dagger a_j + a_i a_j^\dagger)},
\end{aligned}
\end{equation}
where the subscripts $1,2,-,+$ correspond to the aforementioned elements, $\omega_i$ and $\alpha_i$ are their frequencies and anharmonicities, and the $g_{ij}$ are the coupling rates between them. In our particular implementation, $\alpha_+=0$ for the bus cavity. In this Letter, we present data measured on two separate devices. In device A (B), the qubits are fixed-frequency transmons with $\omega_{1}/2\pi=4.973\,(6.143)$ GHz and $\omega_{2}/2\pi=5.163\,(6.421)$ GHz, the bus cavity has a resonance frequency of $\omega_+/2\pi=$7.036 (7.073) GHz, and a maximum tunable coupler frequnecy $\omega_-^\mathrm{max}/2\pi= 7.180\,(7.191)$.  $\omega_-$ can be tuned via an on-chip bias line which changes the flux through the coupler SQUID loop.  The device parameters were obtained by fitting the two-tone spectroscopy measurements with the eigenenergies obtained from Eqn.~\eqref{eq:TunableCoupler}, the details of which can be found in the supplementary materials \cite{supplement}.

%The tunable coupler is comprised of a SQUID loop, and the bus cavity has resonance frequency of $\omega_+/2\pi=7.036$ GHz and $\alpha_+=0$. The coupler qubit frequency $\omega_-$ can be tuned via \tmh{an on chip} flux bias line. \psm{In device A, }\tmh{the large anharmonicities, $\alpha_{1,2}=400$ MHz, of the qubits were chosen to ensure that $|\omega_1-\omega_2|<\alpha_i$, however, this increases the charge dispersion of the transmons making them more susceptible to charge noise.  A second device, with higher qubit frequencies and an $E_J/E_C$ ratio of 50 was characterized and used for the iSWAP two-qubit measurements.}

Both devices are operated in the dispersive regime, where  $|\omega_i-\omega_j|\gg g_{ij}$, to minimize population leakage into the tunable coupler during gate operations and decoherence induced by flux noise in the coupler qubit~\cite{mckay2016}. In this regime, $\zeta$ can be calculated using fourth-order perturbation theory~\cite{zhu2013}, and the analysis shows that the necessary criterion for zero $\zeta$ and high qubit-qubit coupling strength is that the bus cavity (tunable coupler) be above (below) both qubits in frequency and one qubit be in the straddling regime of the other, i.e.
\begin{equation}\label{eq:zeroZZcriteria}
\omega_-<\omega_{1, 2}<\omega_+,\quad|\omega_1-\omega_2|<\alpha_{1,2}.
\end{equation}
Tunability of $\zeta$ is realized by adjusting the frequency of the tunable coupler, $\omega_-$. Zero $\zeta$ can be achieved when there is a destructive interference between \textit{ZZ} interactions caused by the bus cavity and the tunable coupler. %To ensure the device is in the straddling regime, device A was designed with large qubit anharmonicities, $\alpha_{1,2}=400$ MHz. However for low frequency qubits, this results in an $E_J/E_C\approx20$ making the qubits more susceptible to charge noise.  Device B was designed with an $E_J/E_C\approx50$ to improve qubit coherence.

%\tmhout{We perform two-tone spectroscopy on both computational qubits while tuning the frequency of the coupler qubit via external magnetic field, and the result is shown in \autoref{fig:ZZCalib}(a)}. \tmh{We perform two-tone spectroscopy on both computational qubits while tuning the frequency of the coupler qubit via an on chip flux bias line.}  When the coupler is brought into resonance with either qubit, an avoided crossing is observed and the corresponding coupling rate $g_{1-},\,g_{2-}$ can be extracted by fitting the data to the eigenenergies of Eqn.~\eqref{eq:TunableCoupler} (See supplementary materials for the obtained parameter values).

\begin{figure}[htbp]
    \centering
    \includegraphics{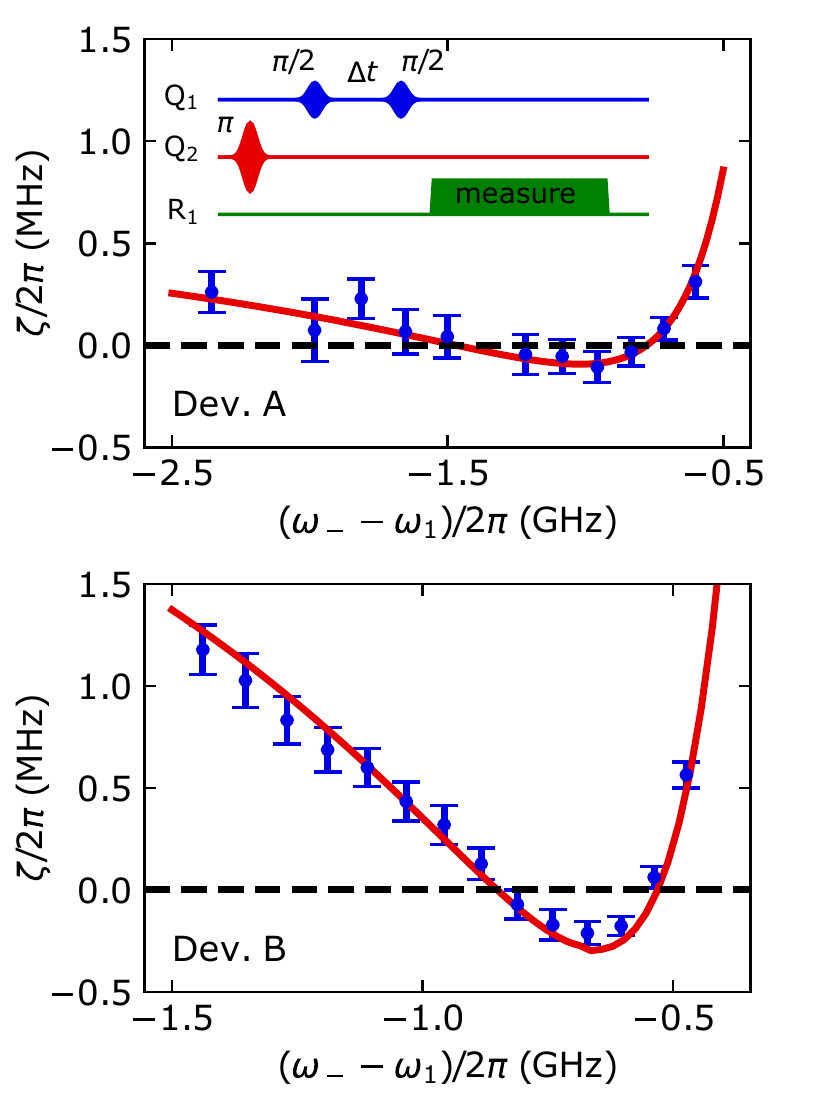}
    \caption{%Calibration of tunable $\zeta$ device. %(a) Spectroscopy of qubits and coupler. Red dashed lines are fit to the eigenenergies of Eqn.~\eqref{eq:TunableCoupler}. Black dashed lines correspond to $\omega_i+g_{i+}^2/\Delta_{i+}\ (i=1, 2)$ and $\omega_-(\Phi)$. (b) 
    \textit{ZZ} interaction strength, $\zeta$, as a function of the tunable coupler detuning from Q$_1$ for device A (top) and device B (bottom). Both devices are in the straddling regime and have two zero $\textit{ZZ}$ points.  The value of $\zeta$ (blue points) is obtained by cross\tmh{-}Ramsey calibration, where the frequency of Q$_1$ is measured with and without a $\pi$ pulse to Q$_2$ at the start of the experiment (illustrated in panel (a) inset). The red line is the theoretical result from fourth-order perturbation theory using the fitted parameters~\cite{supplement}.  The error bars correspond to the fitting error of the Ramsey data.  The strength of $\zeta$ changes more rapidly away from the null point in device B due to the smaller detuning between the fixed coupler and the qubits.}
    \label{fig:ZZCalib}
\end{figure}

The frequency shift of qubit 1 when the state of qubit 2 changes from ground to excited corresponds to the quantity $\zeta=\omega_{|11\rangle}-\omega_{|10\rangle}-\omega_{|01\rangle}$, which represents the ZZ coupling strength between the two qubits. $\zeta$ is measured via a cross-Ramsey measurement which involves measuring the qubit frequency with a Ramsey experiment while initializing the other qubit in either its ground or excited state [see inset of \autoref{fig:ZZCalib}(a) for the pulse sequence]. The dependence of $\zeta$ on the coupler frequency $\omega_-(\Phi)$ is mapped out in \autoref{fig:ZZCalib} via cross-Ramsey measurements on qubit 1. Based on the criterion in Eqn.~\eqref{eq:zeroZZcriteria}, we tune the frequency of the coupler qubit to be below those of both qubits and observe that $\zeta$ crosses zero at $(\omega_--\omega_1)/2\pi=-1.47\  (-0.84)$\ GHz and $-0.75\ (-0.53)$ GHz for device A (B). We note that having two points for which $\zeta=0$ is not universal, and depending on the device parameters (i.e. $\omega_j\ \mathrm{and}\ g_{ij}$) there can also be one or zero solutions.%The theoretical curves in \autoref{fig:ZZCalib} are calculated from the numerical diagonalization of the Hamiltonian in Eqn.~\eqref{eq:TunableCoupler} using the parameters obtained from fitting the spectroscopy data \tmh{(See supplementary materials)}. 

\begin{figure}[htbp]
    \centering
    \includegraphics{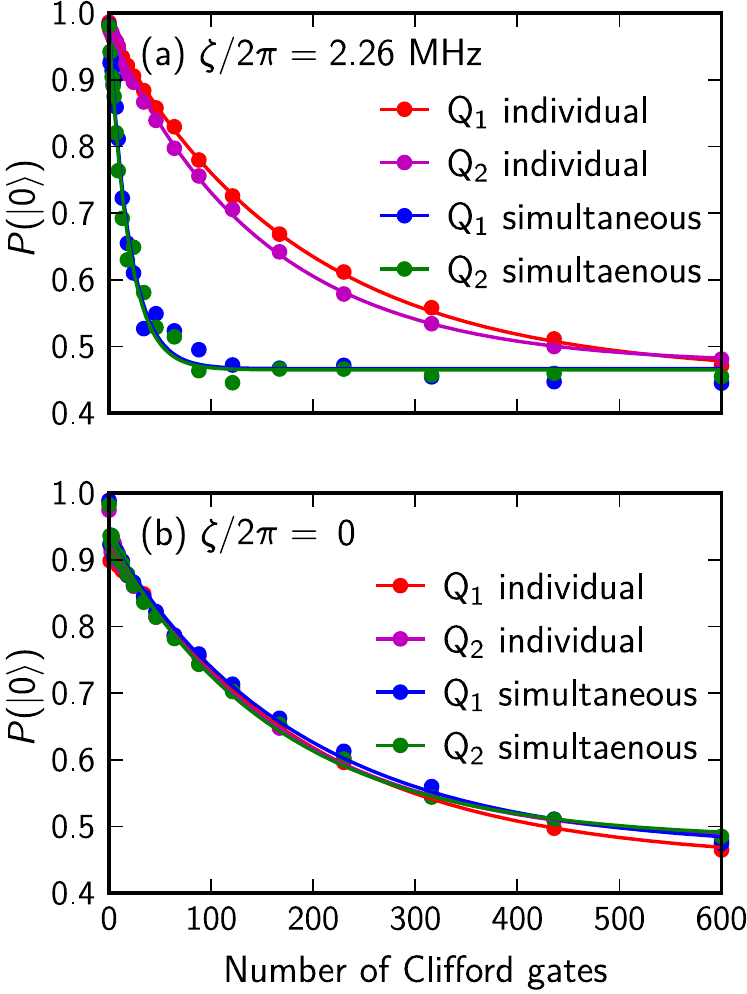}
    \caption{Qubit crosstalk characterization of device A. Individual and simultaneous RB are performed on Q$_1$ and Q$_2$, when the tunable coupler is tuned to give (a) large and (b) small $|\zeta|$. Red and magenta curves correspond to the individual RB measurements and exhibit a primary gate fidelity of $F_I > 99.8\,\%$ irrespective of the magnitude of $\zeta$. The blue and green curves represent the simultaneous RB measurements which demonstrate strong dependence of the primary gate fidelity $F_S$ on the \textit{ZZ} crosstalk. For small $\zeta$, the difference $F_I-F_S$ is within $0.01\,\%$, while for large $\zeta$, the difference is more than $1.15\,\%$.}
    \label{fig:SimRB}
\end{figure}

To further characterize the effect of $\zeta$ on qubit crosstalk, we utilize the simultaneous RB protocol, where the difference in gate fidelity between individual ($F_I$) and simultaneous ($F_S$) RB provides a figure of merit for addressability and crosstalk~\cite{gambetta2012}. The pulses used for single-qubit gates have Gaussian envelopes truncated at $4\sigma$ in total, with $\sigma=6.4$ ns. Derivative removal via adiabatic gate (DRAG)~\cite{motzoi2009,chow2010} is used for pulse correction reducing phase error and leakage to higher transmon levels. As shown in \autoref{fig:SimRB}, the average gate fidelity, obtained from an exponential fit, for individual RB is $F_I > 99.8\,\% $ for the primary gate set $\{\mathbb{I},\,X_{\pm\pi/2},\,Y_{\pm\pi/2},\,X_\pi,\,Y_\pi\}$ for both qubits, which is consistent with the coherence-limited fidelity of $99.81\,\%$ estimated from the device A parameters of $T_1=[15.2\ \mu\text{s},12.1\ \mu\text{s}]$ and $T_2=[4.2\ \mu\text{s},4.0\ \mu\text{s}]$.  The individual RB fidelity is not affected by the magnitude of $\zeta$ whereas the gate fidelity from simultaneous RB decreases with increasing $\zeta$. When $\zeta/2\pi = 0$, $F_I-F_S$ is less than $0.01\,\%$, indicating that crosstalk is suppressed to a level below the gate error for this device. By contrast, when $\zeta/2\pi=2.26$ MHz the gate error increases by an order of magnitude, and \textit{ZZ} crosstalk becomes the dominant source of gate error. We find good agreement between these results and numerical calculation using a Kraus map model for RB \cite{mckay2017,johansson2013}.

After characterizing single qubit gates, we now exhibit two-qubit entangling interactions to establish a universal quantum gate set in this architecture.  We present the two-qubit gate results from device B, which has improved coherence times  ($T_2=[22.5\ \mu\text{s}, 9.3\ \mu\text{s}]$) over device A, due to its larger $E_J/E_C$ ratio (53 compared to 20). The two-qubit gate is implemented using parametric modulation of the tunable coupler~\cite{dewes2012,mckay2016}. When the magnetic flux threading the SQUID loop of the tunable coupler is modulated around $\Phi=\Theta$ at frequency $\omega_\Phi=\omega_2-\omega_1$, phase $\phi$ and amplitude $\delta$, i.e. $\Phi(t)=\Theta+\delta\cos(\omega_\Phi t+\phi)$,
the effective exchange coupling between the two qubits in their rotating frame is
\begin{equation}\label{eq:HiSWAP}
H_{\text{int}}/\hbar=\frac{\delta}{2}\frac{\partial J}{\partial\Phi}\left(a_1^\dagger a_2 e^{-i\phi}+a_1 a_2^\dagger e^{i\phi}\right),
\end{equation}
where
\begin{equation}\label{eq:Jexpression}
J = \sum_{j=\pm}\frac{g_{1j}g_{2j}}{2}\left(\frac{1}{\omega_1-\omega_j}+\frac{1}{\omega_2-\omega_j}\right)
\end{equation}
is the effective exchange interaction mediated by the couplers. The parametric modulation brings the computational qubits effectively into resonance and can be used to implement an iSWAP gate. Importantly, the effective coupling strength depends on the derivative of $J$ with respect to $\Phi$, and in this device, despite small $\zeta$, $\delta\cdot\partial J/\partial\Phi$ can be tuned from zero to a few MHz for moderate modulation amplitude $\delta$. An efficient two-qubit gate can therefore be implemented while \textit{ZZ} crosstalk is suppressed. 
%\tmh{We perform a two qubit parametric gate in Device B, as it has a larger $\partial J/\partial\Phi$ and therefore faster gate for a smaller drive amplitude.  Additionally, the tunable coupler frequency is at the $\zeta=0$ points is higher in frequency compared to device A, which will minimize thermal occupation of the coupler (See supplemental information).} 

\begin{figure*}[htbp]
    \centering
    \includegraphics[scale=0.26]{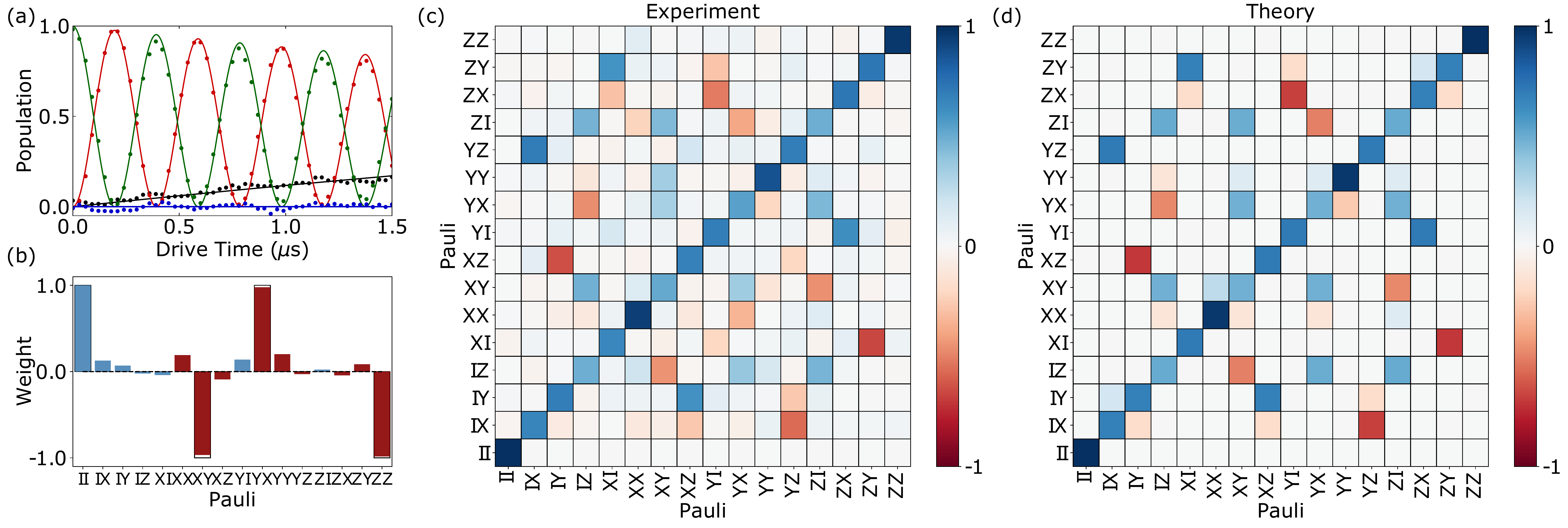}
    \caption{Parametric flux modulation performed on device B. (a) The population of the four two-qubit basis states, $|00\rangle$, $|01\rangle$, $|10\rangle$, and $|11\rangle$ (black, red, green, and blue respectively), for a magnetic flux modulation frequency $\omega_\Phi/2\pi=(\widetilde{\omega}_2-\widetilde{\omega}_1)/2\pi = 275$ MHz. Maximum population exchange between $Q_1$ and $Q_2$ is achieved when the modulation duration is $190$ ns. (b) The expectation values of the Pauli set of two-qubit operators plotted for a modulation time of $95$ ns.  The single (blue bars) and two-qubit  (red bars) correlators are shown along with the theoretically expected values (black bounding boxes). (c) Experimental and (d) Theoretical Pauli transfer matrices for the $\sqrt{\mathrm{iSWAP}}$ gate for the 16 different input states and output states .  The theoretical process tomography results include an additional $Z_{\pi/12}$ gate on Q$_2$ to account for the single qubit phase accumulated during the flux modulation pulse.}
    \label{fig:iSWAP}
\end{figure*}

We implement the following pulse scheme -- (i) initialize in the computational state $|10\rangle$ by applying a $X_\pi$ gate on Q$_1$, (ii) apply flux modulation drive to the coupler for varying durations, (iii) measurement of the qubit populations. The modulation frequency is fixed at the detuning of the two qubits (i.e. $\omega_\Phi=\widetilde{\omega}_2-\widetilde{\omega}_1$, where $\widetilde{\omega}_{1,2}$ are the qubit frequencies in the presence of flux modulation). The DC flux bias $\Theta$ is chosen such that $\zeta=0$ based on \textit{ZZ} calibration and simultaneous RB characterization.  The result is shown in \autoref{fig:iSWAP}(a), where flux modulation for a duration of 190 ns leads to maximum population exchange between states $|10\rangle$ and $|01\rangle$.

To perform an entangling gate between the two qubits, we utilize a $\sqrt{\mathrm{iSWAP}}$ gate, where the flux modulation is turned on for 95 ns.  For this modulation time, the two-qubit system forms the maximally entangled $1/\sqrt{2}\left(|10\rangle+i|01\rangle\right)$ state (up to a single qubit phase rotation of $Z_{\pi/12}$ on Q$_2$ acquired when the modulation drive is turned on).  We perform quantum state tomography of the prepared state, shown in~\autoref{fig:iSWAP}(b), and obtain a raw state fidelity of 98.5\%. Measurement noise and imperfect readout fidelity can lead to an unphysical density matrix. Therefore, we calculate the nearest positive semi-definite density matrix with unit trace, $\rho_p$, by minimizing its Hilbert-Schmidt distance, ie. $D(\rho_\mathrm{p},\rho_\mathrm{m}) = \mathrm{Tr}(\rho_\mathrm{p}-\rho_\mathrm{m})^2$~\cite{vedral1998}, with the measured density matrix, $\rho_\mathrm{m}$, resulting in a fitted state fidelity of 99.4\% with a concurrence of 99\% and $D(\rho_\mathrm{p},\rho_\mathrm{m}) = 0.004$.
%\tmh{This fidelity is limited by the T$_1$/T$_2$ of each qubit?}

We perform quantum process tomography \cite{nielsen2010} by implementing the $\sqrt{\mathrm{iSWAP}}$ on 16 independent two qubit input states and construct the Pauli transfer matix, $\mathcal{R}$, which is shown in~\autoref{fig:iSWAP}(c,d). The gate fidelity can be determined from the $\mathcal{R}$ map through the expression $F_g=(\mathrm{Tr}[\mathcal{R}_\mathrm{id}^\dagger\mathcal{R}_\mathrm{exp}]+2n)/(4n^2+2n)$ where $\mathcal{R}_\mathrm{id}$ and $\mathcal{R}_\mathrm{exp}$ are the ideal and experimental $\mathcal{R}$ maps and $n$ is the number of qubits~\cite{Chow2012}.  We extract a gate fidelity of $\mathcal{F}_g(\mathrm{raw})=96.3\%\ \mathrm{and}\ \mathcal{F}_g(\mathrm{fit})=94.8\%$ with a nonphysical error of $\gamma_{np}=0.5||\mathcal{R}_\mathrm{raw}-\mathcal{R}_\mathrm{fit} ||_2/(2n)=0.055$. The gate fidelity is close to the coherence limit of 98.4\%. This discrepancy in the fidelity is attributed to state preparation and measurement (SPAM) errors, and to population leakage out of the computational subspace, as previously reported~\cite{dewes2012,mckay2016}. 

%\tmhout{In the experiment we find a noticeable amount of thermal population in the coupler qubit, evidenced by a splitting of frequency in both qubits' spectrum near the avoided crossing with the coupler qubit. The thermally dressed frequency for qubit 1 can be seen in the measured spectroscopy, and comparison between numerical simulation and measured data results in estimated effective coupler temperature of $\sim 90$ mK (See supplementary materials for details). The thermal excitation in the coupler not only changes the detuning $\widetilde{\omega}_1-\widetilde{\omega}_2$ but also changes the strength of the iSWAP operation. This results in $\sim 10\,\%$ of the qubit population remaining unswapped during the iSWAP operation, as seen in \autoref{fig:iSWAP}(a). Although the effect of flux noise on the fidelity of a flux-modulated parametric gate has been studied before~\cite{didier2018}, infidelity due to thermal population of the coupler has not previously been explored. We find that effective coupler temperature plays a crucial role in determining the fidelity of the iSWAP gate as shown in \autoref{fig:iSWAP}(b). Thermalization of the coupler to the base temperature of the dilution refrigerator ($\sim 15$mK) will enable high fidelity (infidelity $<10^{-5}$) iSWAP gates. Better thermalization techniques~\cite{yeh2017}, post selection methods and quantum optimal control schemes~\cite{machnes2018,malis2018} can be used to further optimize the iSWAP gate.}

In conclusion, we have demonstrated a way to achieve zero \textit{ZZ} crosstalk by utilizing quantum interference in a tunable coupler device. This device allows us to operate in an optimal configuration to suppress qubit crosstalk, and the tunable \textit{ZZ} interaction strength provides a useful tool to study the impact of crosstalk in cQED systems. $\sqrt{\mathrm{iSWAP}}$ gate was performed while maintaining zero \textit{ZZ} crosstalk. This architecture paves the way for crosstalk free multiqubit quantum processors. As the parameter regime for achieving zero \textit{ZZ} is similar to that of a cross-resonance gate \cite{chow2011,magesan2018}, implementing this form of two qubit gate is a natural extension of this device architecture.

We note that a recent theoretical architecture similar to our work has been independently proposed in~\cite{yan2018} for achieving zero qubit-qubit dipole coupling through quantum interference. 

%\tmh{\psmout{For two qubit state assignment, we calibrate both individual and joint qubit readout.}\psm{**I think joint qubit readout means something else**}  For individual qubit readout, we measure the result of $\sim10^4$ single shot measurements with either an $X_\pi$ or $\mathbb{I}$ to prepare either $|1\rangle$ or $|0\rangle$ on each qubit separately, histograming the resulting measurements and setting a homodyne voltage threshold which maximizes visibility between the two states.  An additional two qubit state correction is made by preparing and reading out the four input states $|00\rangle$,$|01\rangle$,$|10\rangle$,and $|11\rangle$.}

%\psm{Description of readout and other important experimental details. single qubit rotation after the swap or root swap to get all phases ok.}  

This work is supported by IARPA under contract W911NF-10-1-0324.
\bibliography{references}

%%%%%%%%%% 

%%note = {Page 394, exercise 8.34}, for nielsen and chuang
%\include{supplement}
\end{document}

\title{Supplementary materials for ``Suppression of Qubit Crosstalk\\ in a Tunable Coupling Superconducting Circuit''}
\author{Pranav Mundada}\thanks{These authors contributed equally to this work.}
\author{Gengyan Zhang}\thanks{These authors contributed equally to this work.}
\author{Thomas Hazard}\thanks{These authors contributed equally to this work.}
\author{Andrew Houck}
\email{aahouck@princeton.edu}
\affiliation{Department of Electrical Engineering, Princeton University, Princeton, New Jersey 08540, USA}
\maketitle{}

\section{Device parameters}
The coupler frequency in a full flux quantum is measured using two-tone spectroscopy and the bus cavity frequency is measured by monitoring the ac Stark shift of either qubit while sweeping the frequency of a cavity populating tone. The coupling parameters are obtained by fitting the spectroscopy data in each device, as shown in figure S1.

\begin{table}[htbp]
    \centering
    \caption[Tunable $\zeta$ device parameters]{Tunable $\zeta$ device parameters.}\label{tab:ZZParams}
    \begin{tabular}{cccc}
        \toprule
        Parameter & Symbol & Dev. A  & Dev. B\\
        \hline
        Qubit 1 frequency & $\omega_1/2\pi$ & $4.973$ GHz & $6.143$ GHz\\
        Qubit 1 anharmonicity & $\alpha_1/2\pi$ & $400$ MHz & $330$ MHz \\
        Qubit 1 relaxation time & $T_1^{(1)}$ & $15.2$ $\mu$s & $12.5$ $\mu$s\\
        Qubit 1 coherence time & $T_{2\text{E}}^{(1)}$ & 4.2 $\mu$s & $22.5$ $\mu$s\\
        Qubit 2 frequency & $\omega_2/2\pi$ & $5.163$ GHz & $6.421$ GHz\\
        Qubit 2 anharmonicity & $\alpha_2/2\pi$ & $400$ MHz & $330$ MHz\\
        Qubit 2 relaxation time & $T_1^{(2)}$ & $12.1$ $\mu$s & $7.0$ $\mu$s\\
        Qubit 2 coherence time & $T_{2\text{E}}^{(2)}$ & $4.0$ $\mu$s & $9.3$ $\mu$s\\
        Bus cavity frequency & $\omega_+/2\pi$ & $7.036$ GHz & $7.073$ GHz\\
        Maximum coupler frequency & $\omega_-^{\max}/2\pi$ & $7.18$ GHz & $7.19$ GHz\\
        Coupler anharmonicity & $\alpha_-/2\pi$ & $750$ MHz & $290$ MHz\\
        (Qubit 1, bus cavity) coupling & $g_{1+}/2\pi$ & $135$ MHz & $102$ MHz\\
        (Qubit 2, bus cavity) coupling & $g_{2+}/2\pi$ & $135$ MHz & $102$ MHz\\
        (Qubit 1, coupler) coupling & $g_{1-}/2\pi$ & $95$ MHz & $85$ MHz\\
        (Qubit 2, coupler) coupling & $g_{2-}/2\pi$ & $95$ MHz & $85$ MHz\\
        \toprule
    \end{tabular}
\end{table}
\begin{figure}[htbp]
    \centering
    \includegraphics{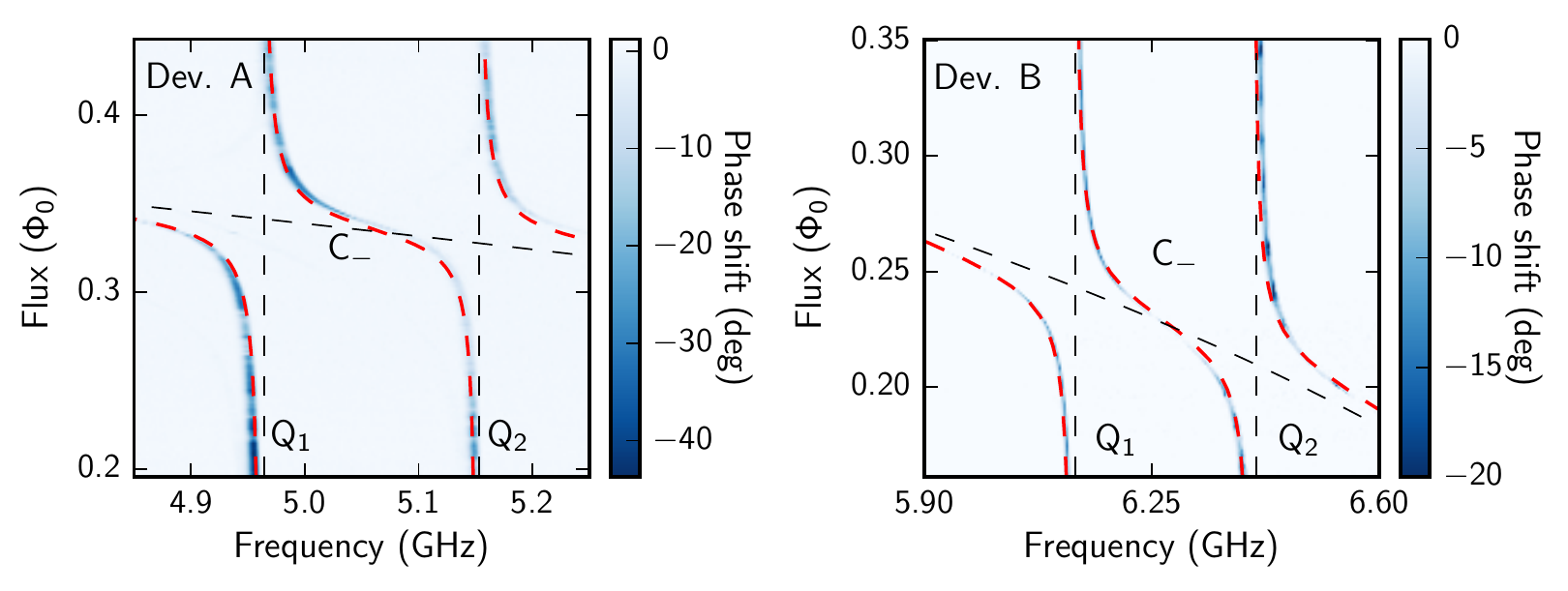}
    %"C:\Users\Tom Hazard\Desktop\python\untitled0.py" for thing about 
    \caption{Two-tone spectroscopy data for devices A (left) and B (right).  The combined phase response for each readout resonator plotted as a function of the spectroscopic tone frequency and flux through the tunable coupler.  The red dashed lines are a found by numerical diagonalization of~\autoref{eq:zetaH}. Black dashed lines indicate bare qubit and coupler frequencies.}
    \label{fig:Spec}
\end{figure}
\section{RB simulations}
For the simulation of RB sequences, we follow the protocol used in~\cite{mckay2017}. For ease of reading, we describe the protocol here using the same notation as that used in~\cite{mckay2017}. The accrued error is measured by tracking the density matrix as we go through the sequence of gates after starting in the ground state. For each gate in the RB sequence, we first apply an ideal gate unitary transformation, followed by a \textit{ZZ} unitary transformation and a density matrix map to account for decoherence. The maps used are
\begin{equation}
\Lambda_\text{gate}[\rho] = U_g \cdot \rho \cdot U_g^\dagger,
\end{equation}

\begin{equation}
\Lambda_{ZZ}[\rho] = U_{ZZ} \cdot \rho \cdot U_{ZZ}^\dagger, 
\end{equation}

\begin{equation}
\Lambda_{T_1, T_2}[\rho] = \frac{1-e^{-t/T_2}}{2} \pmb{Z} \cdot \rho \cdot \pmb{Z} + \frac{1+e^{-t/T_2}}{2} \rho + (1-e^{-t/T_1})|0\rangle\langle 1|\cdot \rho \cdot |1\rangle\langle 0| - (1-e^{-t/T_1})|1\rangle\langle 1|\cdot \rho \cdot |1\rangle\langle 1|,
\end{equation}
where $\pmb{Z}$ is the Pauli-Z operator.

The update to the density matrix after each gate can be expressed as follows
\begin{equation}
\rho_{t+1} = \Lambda_{T_1,T_2,\text{Q}1} \circ \Lambda_{T_1,T_2,\text{Q}2} \circ \Lambda_{ZZ} \circ \Lambda_\text{gate,Q1}\circ \Lambda_\text{gate,Q2}[\rho_t].
\end{equation}
The gate duration used for simulation is 22 ns and the coherence values used are for device A, $T_1= [15.2\ \mu s,\ 12.1\ \mu s]$ and $T_2=[4.2\ \mu s,\ 4\ \mu s]$. The measured fidelity matches with the simulation results.
\begin{figure}[htbp]
    \centering
    \includegraphics{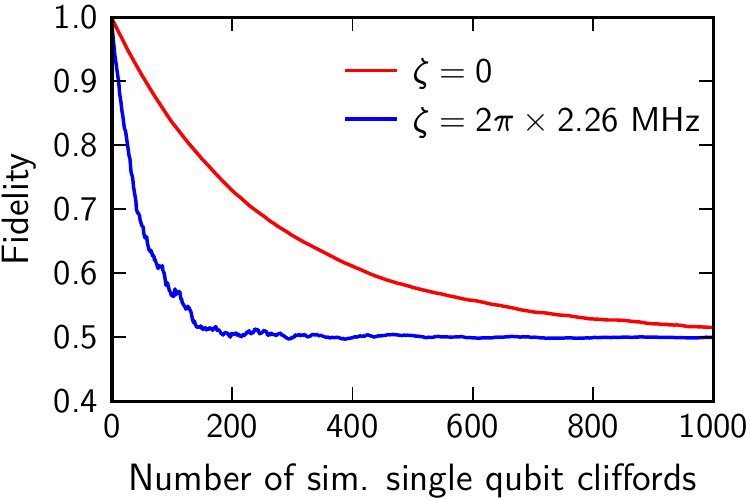}
    \caption{Kraus map model based simulation of simultaneous randomized benchmarking schemes gives fidelity of $F_S = 99.8\,\%$ for $\zeta = 0$ and $F_S = 98.5\,\%$ for $\zeta = 2\pi\times 2.26$ MHz. This matches well with the experimental results shown in the main text.}
    \label{fig:RBFit}
\end{figure}

\section{$\zeta$ calculation}
We start from the Hamiltonian in Eqn.(1) in the main text,
\begin{equation}\label{eq:zetaH}
\begin{aligned}
H &= H_0 + V\\
  &=\sum_{i=1,2,\pm}{\hbar\left(\omega_i a_i^\dagger a_i - \frac{\alpha_i}{2}a_i^\dagger a_i^\dagger a_i a_i\right)} + \sum_{\substack{i=1,2\\j=\pm}}\hbar{g_{ij}(a_i^\dagger a_j + a_i a_j^\dagger)}.
\end{aligned}
\end{equation}
and denote the eigenstates and eigenfrequencies by $|n_1 n_2 n_+ n_-\rangle$ and $\omega_{n_1 n_2 n_+ n_-}$. The detunings $\Delta_{ij}$ are the differences between unperturbed, single-excitation energy levels, e.g., $\Delta_{1+} = \omega_{1000}^{(0)}-\omega_{0010}^{(0)}$, etc. The ZZ coupling rate $\zeta$ between qubit 1 and 2 (assuming the couplers are in their ground states) is given by
\begin{equation}\label{eq:zetaDef}
\zeta = \omega_{1100} - \omega_{1000} - \omega_{0100}\quad(\omega_{0000} = 0\text{ for all orders}).
\end{equation}

We use fourth order perturbation theory outlined in~\cite{zhu2013},
and the expression for $\zeta$ is
\begin{equation}\label{eq:zetaResult}
\begin{aligned}
\zeta&=\omega_{1100}^{(4)} - \omega_{1000}^{(4)} - \omega_{0100}^{(4)}\\
&=\frac{2 g_{1+}^2 g_{2+}^2}{\Delta_{1+}+\Delta_{2+}+\alpha_+}\left(\frac{1}{\Delta_{1+}}+\frac{1}{\Delta_{2+}}\right)^2 + \frac{2 g_{1-}^2 g_{2-}^2}{\Delta_{1-}+\Delta_{2-}+\alpha_-}\left(\frac{1}{\Delta_{1-}}+\frac{1}{\Delta_{2-}}\right)^2\\
&\quad +\left(\frac{g_{1+}g_{2+}}{\Delta_{1+}}+\frac{g_{1-}g_{2-}}{\Delta_{1-}}\right)^2\left(\frac{2}{\Delta_{12}+\alpha_2}-\frac{1}{\Delta_{12}}\right)\\
&\quad +\left(\frac{g_{1+}g_{2+}}{\Delta_{2+}}+\frac{g_{1-}g_{2-}}{\Delta_{2-}}\right)^2\left(\frac{2}
{\Delta_{21}+\alpha_1}-\frac{1}{\Delta_{21}}\right)\\
&\quad + \left[g_{1+}g_{2-}\left(\frac{1}{\Delta_{1+}}+\frac{1}{\Delta_{2-}}\right)+g_{1-}g_{2+}\left(\frac{1}{\Delta_{1-}}+\frac{1}{\Delta_{2+}}\right)\right]^2\frac{1}{\Delta_{1+}+\Delta_{2-}}\\
&\quad - \left(\frac{g_{1+}^2}{\Delta_{1+}^2}+\frac{g_{1-}^2}{\Delta_{1-}^2}\right)\left(\frac{g_{2+}^2}{\Delta_{2+}}+\frac{g_{2-}^2}{\Delta_{2-}}\right) - \left(\frac{g_{2+}^2}{\Delta_{2+}^2}+\frac{g_{2-}^2}{\Delta_{2-}^2}\right)\left(\frac{g_{1+}^2}{\Delta_{1+}}+\frac{g_{1-}^2}{\Delta_{1-}}\right).
\end{aligned}
\end{equation}

\begin{figure}[htbp]
	\centering
	\includegraphics{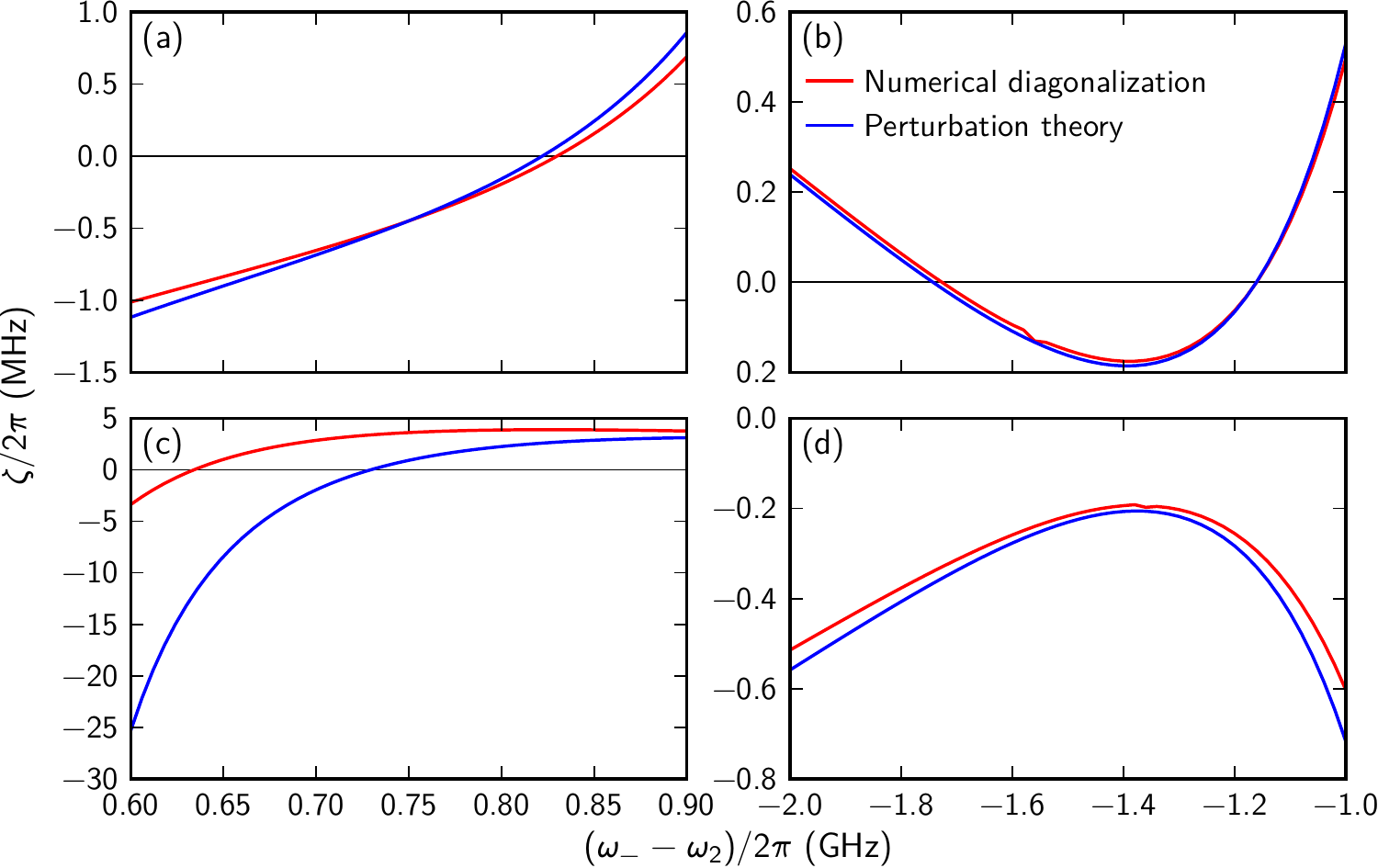}
	\caption{\textit{ZZ} coupling rate calculated from perturbation theory (blue) and numerical diagonalization (red). (a) Qubit frequencies are far apart with one coupler in between. (b) Qubits are in straddling regime, with one coupler above and one below the qubits in frequency. (c) The qubits are in straddling regime with one coupler above the qubit frequencies. (d) Qubits are out of straddling regime with one coupler above and one coupler below the qubits in frequency. The parameters used for each configuration are listed in \autoref{tab:zetaParams}.}
	\label{fig:ZZTheory}
\end{figure}

To show the possibility of zero \textit{ZZ} interaction, we calculate $\zeta$ for different parameter configurations using Eqns.~\eqref{eq:zetaResult} and \eqref{eq:zetaDef}, and the results are shown in the \autoref{fig:ZZTheory}. The parameter configurations for each plot are listed in \autoref{tab:zetaParams}.

\begin{table}[htbp]
    \centering
    \caption{Parameter configurations for \textit{ZZ} coupling rate calculation in \autoref{fig:ZZTheory}.}\label{tab:zetaParams}
    \begin{tabular}{ccc}
        \toprule
        Figure & Configuration & Parameters ($2\pi\cdot$MHz)\\
        \hline
        (a) & Qubits far apart, & $\alpha_1=\alpha_2=350,\ \alpha_-=750,$ \\
        & one coupler in between. & $\Delta_{12}=1500,\  g_{1-}=g_{2-}=140.$ \\[1ex]
        (b) & Qubits in straddling regime,& $\alpha_1=\alpha_2=350,\ \alpha_-=750,\ \alpha_+=0,$ \\
        & one coupler above, & $\Delta_{12}=250,\ \Delta_{+2}=1800,$ \\
        & one coupler below. & $g_{1+}=g_{2+}=160,\ g_{1-}=g_{2-}=140.$ \\[1ex]
        (c) & Qubits in straddling regime, & $\alpha_1=\alpha_2=350,\ \alpha_+ = 750,$ \\
        & one coupler above. & $\Delta_{12}=250,\ g_{1+}=g_{2+}=120.$ \\[1ex]
        (d) & Qubits out of straddling regime,& $\alpha_1=\alpha_2=350,\ \alpha_-=750,\ \alpha_+=0,$ \\
        & one coupler above, & $\Delta_{12}=450,\ \Delta_{+2}=1800,$ \\
        & one coupler below. & $g_{1+}=g_{2+}=160,\ g_{1-}=g_{2-}=140.$ \\
        \toprule
    \end{tabular}
\end{table}

From \autoref{fig:ZZTheory} we find that there are several configurations that result in zero $\zeta$. We choose the configuration in (b) because the two qubits are close to each other in frequency and have stronger interaction strength compared to (a), which can potentially lead to fast two qubit gates in addition to zero $\zeta$. In configuration (c) zero $\zeta$ happens at relatively small detuning $\Delta_{-2}/2\pi=634$ MHz, which increases the susceptibility of the qubits to flux noise in the coupler.
\section{Readout calibration}
For the data presented in Fig. 4 of the main text, we perform single shot readout of the two qubits.  Following the method detailed in \cite{dewes2012}, we calibrate the readout for each qubit individually, and the two qubits simultaneously.  For each qubit, we prepare the states $|0\rangle$ and $|1\rangle$ and measure the result to obtain the single qubit readout matrix
\begin{equation}\label{eq:readoutMatrix}
C_i=
\begin{pmatrix}
1-|0\rangle_\mathrm{err} & |1\rangle_\mathrm{err}\\
|0\rangle_\mathrm{err} & 1-|1\rangle_\mathrm{err}
\end{pmatrix},\ \mathrm{for}\ i = 1,2
\end{equation}
where $|0\rangle_\mathrm{err}$ and $|1\rangle_\mathrm{err}$ are the readout error for the states $|0\rangle$ and $|1\rangle$ respectively.  This process is repeated for the four two qubit states, $|00\rangle$,$|01\rangle$,$|10\rangle$, and $|11\rangle$ to obtain the two qubit ``crosstalk" readout matrix, $C_{CT}$.  The final readout correction matrix is obtained by taking the product of the crosstalk matrix with the tensor product of the individual qubit readout matrices, $C_{CT}\cdot\left(C_1 \otimes C_2\right)$ . The raw data is corrected by multiplying the measured state vector by the inverse of the readout correction matrix.

\section{Simulation of \lowercase{i}SWAP fidelity vs Temperature}

Here, we comment on the potential effects of coupler temperature on the two-qubit gate fidelity. From second order perturbation theory, we have the following effective iSWAP coupling strengths for different states of the tunable coupler --

(Tunable coupler population = 0)
\begin{equation*}
J_0=\frac{1}{2}\left[g_{1+}g_{2+}\left(\frac{1}{\Delta_{1+}}+\frac{1}{\Delta_{2+}}\right)+g_{1-}g_{2-}\left(\frac{1}{\Delta_{1-}}+\frac{1}{\Delta_{2-}}\right)\right],
\end{equation*}

\begin{equation}
\partial J_0/\partial\Phi=\frac{1}{2}\left[g_{1-}g_{2-}\left(\frac{1}{\Delta_{1-}^2}+\frac{1}{\Delta_{2-}^2}\right)\right] \partial\omega_-/\partial\Phi.
\end{equation}

(Tunable coupler population = 1)
\begin{equation*}
J_1=\frac{1}{2}\left[g_{1+}g_{2+}\left(\frac{1}{\Delta_{1+}}+\frac{1}{\Delta_{2+}}\right)+2g_{1-}g_{2-}\left(\frac{1}{\Delta_{1-}+\alpha_-}+\frac{1}{\Delta_{2-}+\alpha_-}\right)-g_{1-}g_{2-}\left(\frac{1}{\Delta_{1-}}+\frac{1}{\Delta_{2-}}\right)\right],
\end{equation*}

\begin{equation}
\partial J_1/\partial\Phi=\frac{1}{2}\left[2g_{1-}g_{2-}\left(\frac{1}{(\Delta_{1-}+\alpha_-)^2}+\frac{1}{(\Delta_{2-}+\alpha_-)^2}\right)-g_{1-}g_{2-}\left(\frac{1}{\Delta_{1-}^2}+\frac{1}{\Delta_{2-}^2}\right)\right]\partial\omega_-/\partial\Phi.
\label{eqn:J1}
\end{equation}

As seen from the last two terms in Eqn.~\eqref{eqn:J1}, the effective iSWAP strength is decreased for excited coupler due to destructive intereference. For our device A (B) parameters, we have $\left|\frac{\partial J_0/\partial\Phi}{\partial J_1/\partial\Phi}\right| = 3.2\ (6.1)$. Note that the resonance condition for an iSWAP gate is independent of the coupler population. Since we calibrate the duration of flux modulation to get an $\sqrt{\mathrm{iSWAP}}$ gate for the coupler in ground state, we model the unitary for the excited coupler as a partial $\sqrt{\mathrm{iSWAP}}$ gate. In the simulation of $\sqrt{\mathrm{iSWAP}}$ gate fidelity (for a 95 ns long gate) with finite temperature, we use the following map
\begin{equation}
\Lambda_\text{FM}[\rho] = (1-p)\ U_{\sqrt{\mathrm{iSWAP}}} \cdot \rho \cdot U_{\sqrt{\mathrm{iSWAP}}}^\dagger \ +\ p\ U_\text{FM,1} \cdot \rho \cdot U_\text{FM,1}^\dagger,
\end{equation}
where $p$ is the thermal population in the tunable coupler and $U_\text{FM,1}$ is the effective unitary due to flux modulation with the coupler excited,
\begin{equation*}
U_\text{FM,1} = U_{\sqrt{\mathrm{iSWAP}}}^{1/\alpha},
\end{equation*}
where $\alpha$ for device A (B) is 3 (6). The $\sqrt{\mathrm{iSWAP}}$ unitary is given by

\begin{equation*}
U_{\sqrt{\mathrm{iSWAP}}} = 
\begin{pmatrix}
1&0 &0 &0 \\
0&1/\sqrt{2} & i/\sqrt{2} &0 \\
0&i/\sqrt{2} & 1/\sqrt{2} &0\\
0& 0& 0&1
\end{pmatrix}
.
\end{equation*}

The update to the density matrix due to the flux modulation for an $\sqrt{\mathrm{iSWAP}}$ gate can be expressed as follows
\begin{equation}
\rho_{f} = \Lambda_{T_1,T_2,\text{Q}1} \circ \Lambda_{T_1,T_2,\text{Q}2} \circ \Lambda_\text{FM}[\rho_0].
\end{equation}

We average the fidelity obtained by flux modulation over 16 different density matrices which form a good basis for two qubit process tomography~\cite{nielsen2010}.
\begin{figure}[htbp]
    \centering
    \includegraphics[scale=1]{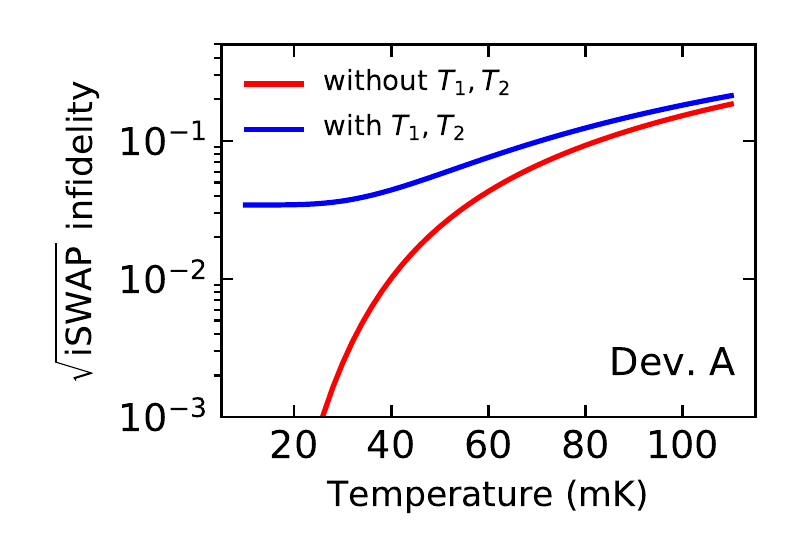}
    \includegraphics[scale=1]{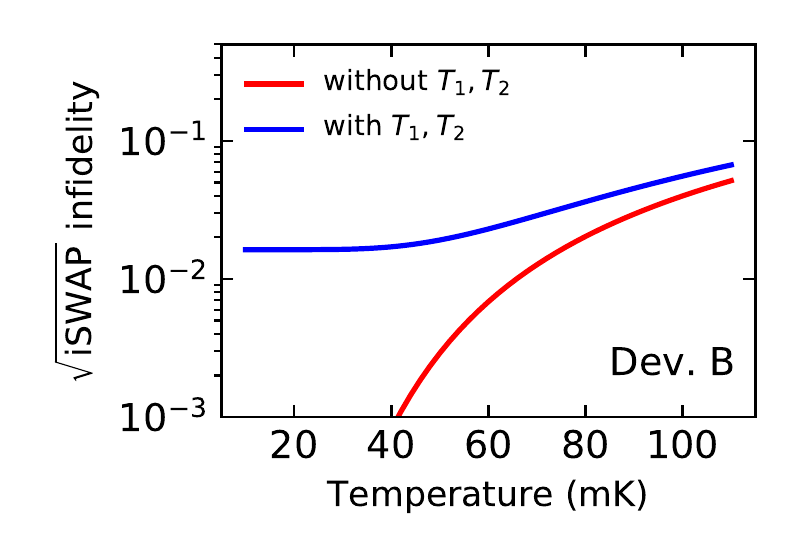}
    \caption{$\sqrt{\mathrm{iSWAP}}$ fidelity for device A (left) and device B (right).  The expected temperature dependent fidelity is higher in device B for all temperatures as $\omega_-$ for the $\zeta=0$ points occur at higher frequency compared to device A.  Device B was measured in a different measurement setup with additional cold attenuation on the cavity and coupler lines, as well as the addition of a K \& L 12 GHz low-pass filter.}
    \label{fig:ZZSpurious}
\end{figure}

\bibliography{references}